\documentclass[12pt]{iopart}
\usepackage{graphicx}
\expandafter\let\csname equation*\endcsname\relax
\expandafter\let\csname endequation*\endcsname\relax
\usepackage{amsmath}
\usepackage{subcaption}
\usepackage{hyperref}
\usepackage{xcolor}

\begin{document}

\title[Improving Performance of Quantum Heat Engines using modified Otto cycle]{Improving Performance of Quantum Heat Engines using modified Otto cycle}

\author{Revathy B. S $^1$, Harsh Sharma $^2$ and Uma Divakaran $^3$}

\address{$^1$ Raman Research Institute, Bengaluru, Karnataka, 560080, India}
\address{$^2$ Department of Physics,  Indian Institute of Technology Bombay, Powai, Mumbai 400076, India}
\address{$^3$ Department of Physics,  Indian Institute of Technology Palakkad, Palakkad, 678557, India}
\ead{revathy@rrimail.rri.res.in, uma@iitpkd.ac.in}
\vspace{10pt}
\begin{indented}
\item[]December 2023
\end{indented}

\begin{abstract}
The efficiency of a quantum heat engine is maximum when the unitary strokes of the quantum Otto cycle are adiabatic.
On the other hand, this may not be always possible due to small energy gaps in the system, especially at the critical point
	where the gap  {between the ground state and the first excited state} vanishes and the system gets excited. With the aim to regain this lost adiabaticity, we modify one 
	of the unitary strokes of the Otto cycle by allowing the system to first evolve with a time dependent Hamiltonian as in the case of a usual Otto cycle, followed by an additional
	evolution with a different time independent Hamiltonian  so that the system reaches a less excited state. This will help in increasing the magnitude of the heat absorbed 
from the hot bath so that the work output and efficiency of the engine can be increased. We demonstrate this method using an 
	integrable model and a non-integrable model as the working medium {and discuss the generality and limitations of this method}. In the case of a two spin system, the optimal value for the time till which the system needs to be freely evolved is calculated analytically in the adiabatic limit. The results show that implementing this modified unitary stroke significantly improves the work output and efficiency of the engine, especially
 when it crosses the critical point. 
\end{abstract}

%\maketitle

\section{Introduction}

Studies on experimental realization of various quantum devices where the working medium is described by a quantum Hamiltonian, have given a 
large push to an already active area of research \cite{Mukherjee_2021, bhattacharjee21quantum, myers22quantum, e23030353}. 
These devices include quantum refrigerators \cite{Abah_2016, Maslennikov17}, quantum batteries \cite{PhysRevA.97.022106, bhattacharjee21quantum}, quantum transistors \cite{PhysRevE.106.024110} with quantum heat engines (QHE) \cite{gemmer2009quantum, PhysRevE.76.031105} receiving maximum attention.
The QHEs are being explored experimentally using a variety of platforms namely, trapped ions or 
ultracold atoms \cite{rossnagel16a,PhysRevLett.123.080602}, optical cavities \cite{schreiber15observation}, using 
NMR techniques \cite{Peterson18}, nitrogen vacancy centres in diamond   \cite{klatzow19experimental}, etc. Theoretical studies on QHE first concentrated 
on few body systems such as single spin systems \cite{PhysRevE.61.4774, PhysRevE.87.012140, PhysRevA.99.062103}, two spin systems \cite{PhysRevE.83.031135, Campisi_2015, Cakmak_2023, Cakmak2021}, and harmonic oscillators \cite{e19040136, Rezek_2006}.
Recently, the focus has shifted to many body systems as working media (WM), mainly to understand the effect of various interactions present in the many body WM. This included 
effects like collective cooperative effects \cite{Niedenzu_2018, PhysRevLett.124.210603, Jaramillo16, PhysRevLett.131.210401},
super-radiance \cite{Hardal15}, many body localization \cite{PhysRevB.99.024203}, and phase transitions \cite{campisi2016power, Fogarty_2021, 10.1088/1367-2630/ac963b,PhysRevE.96.022143}. The performance of these engines are quantified using parameters like efficiency defined as the ratio of work done by the engine to the heat absorbed, and power defined as the ratio of work done to the total cycle time.

Phase transitions in quantum heat engines have been studied in Refs. \cite{campisi2016power, Fogarty_2021, 10.1088/1367-2630/ac963b,PhysRevE.96.022143}. Ref. \cite{PhysRevResearch.2.043247} reported the universality in the finite time dynamics of quantum heat engines using critical working medium, where we also showed that crossing the critical point
 (CP) leads to generation of excitations which are detrimental to the performance of the 	engines. These excitations can be reduced if the CP is crossed slowly, which essentially implies increasing the cycle time of the engine. But this results in vanishing power output as it is inversely proportional to the cycle time.
In order to increase the efficiency of the engine without sacrificing the power output, many control techniques have been put forward, 
the prominent one being the shortcuts to adiabaticity (STA) \cite{PhysRevLett.109.115703, kolodrubetz17geometry, RevModPhys.91.045001}. STA has been highly instrumental in improving the performance of 
quantum machines \cite{delcampo14, Deng13, Beau16, sels17minimizing, PhysRevE.98.032121, delCampo2018, PhysRevResearch.2.023145}. {It has also been experimentally demonstrated using Fermi gas \cite{Diao_2018, doi:10.1126/sciadv.aar5909}.} At the same time it requires additional control Hamiltonian which may involve long range interactions \cite{PhysRevLett.109.115703, kolodrubetz17geometry},
making the process complicated. Also, there have been studies which involves modification of interactions in the unitary strokes, such as in Ref. \cite{chen2019interaction, PhysRevResearch.5.013088}.

 In Ref. \cite{revathybatheng}, bath engineered quantum engine was proposed to overcome the effects of the excitations by removing those modes during the non-unitary strokes which 
 were reducing the work done. On the other hand, in this paper, we propose a simple way to improve the performance of the engine by modifying the unitary stroke {of the quantum Otto cycle} where the system is 
 evolved with a different time independent Hamiltonian for a certain time. We illustrate this idea using an integrable and a non-integrable WM, and show that such a modification indeed aids in better 
 performance of engines compared to normal finite time engines. 
Our protocol resembles the bang-bang protocols used in optimal control theory \cite{PhysRevA.82.063422, doi:10.1021/acs.jpca.5b06090} and has some resemblance with the accidental shortcuts to adiabaticity technique where the non-adiabatic dynamics can lead to an adiabatic one at some specific values of the control parameter \cite{Beau16}.
 The outline of the paper is as follows:
 In Sec. \ref{secII}, the conventional four stroke quantum Otto cycle is discussed followed by the proposed modification. Sec. \ref{secIV} discusses the transverse Ising model whereas
 the antiferromagnetic transverse Ising model with longitudinal field as the WM is discussed in  Sec. \ref{secV}.
 Finally we conclude in Sec. \ref{secVI}.
 
\section{Modified Quantum Otto cycle}
\label{secII}
The working medium (WM) of the many body quantum Otto cycle is described by the Hamiltonian
\begin{equation}
H(t) = H_0 + \lambda(t) H_1,
\end{equation} 
{such that $H_0$ and $H_1$ do not commute, for reasons described later, and} $\lambda$ is the time dependent parameter that can be varied. The usual four stroke quantum Otto cycle consists of two unitary strokes and two non-unitary strokes as described below.
%
%\begin{figure}[h]
%\includegraphics[clip, trim= 2cm 10cm 1cm 4cm,scale=.4]{qoe.pdf}
%\caption{{\bf Schematic diagram of a quantum Otto cycle with a spin chain as the working medium.} }
%\label{fig_cycle}
%\end{figure}

\begin{itemize}
\item[(i)]\underline{\textbf{A $\rightarrow$ B} (\textit{Non-unitary stroke})}: The WM with parameter $\lambda_{1}$ is connected to the hot bath which is at a temperature $T_{H}$ till  a time $\tau_H$ so that it reaches the thermal state at \textbf{B} given by

\begin{equation}\label{eqn_rhoB}
\rho_B = \frac{e^{-\beta_H H(\lambda_1)}}{Z(\lambda_1)},
\end{equation} 
where $\beta_H = \frac{1}{k_{B} T_H}$, (we have taken $k_B = 1$) and $Z$ is the partition function ($Z = \rm{Tr} (e^{-\beta_H H(\lambda_1)}))$.
Let us represent the energy exchanged in this stroke by $\mathcal{Q}_{in}$. 
\item[(ii)]\underline{ \textbf{B $\rightarrow$ C} (\textit{Unitary stroke})}: The WM is decoupled from the hot bath and  $\lambda$ is changed from $\lambda_{1}$ to $\lambda_{2}$ with a speed $1/ \tau_1$.

This unitary evolution is described by the von-Neumann equation of motion:
\begin{equation}
\frac{d\rho}{dt} = -i [H, \rho].
\end{equation}

\item[(iii)]\underline{\textbf{C $\rightarrow$ D} (\textit{Non-unitary stroke})}: The WM with $\lambda_{2}$ is now connected to the cold bath which is at a temperature $T_{C}$ till $\tau_C$ when it reaches the corresponding thermal state at \textbf{D}

\begin{equation}\label{eqn_rhoD}
\rho_D = \frac{e^{-\beta_C H(\lambda_2)}}{Z(\lambda_2)},
\end{equation} 
where $\beta_C = \frac{1}{k_{B} T_C}$.

The energy exchanged in this stroke is represented by $\mathcal{Q}_{out}$.
\item[(iv)]\underline{\textbf{D $\rightarrow$ A} (\textit{Unitary stroke})}: After decoupling from the cold bath, the parameter is changed back to $\lambda_{1}$ from $\lambda_{2}$
with a speed $1/ \tau_2$.

These are the strokes of a conventional four stroke quantum Otto cycle. The energies at the end of each stroke $i$ is calculated using 
\begin{equation}
\mathcal{E}_{i} = \rm{Tr} (H_i \rho_i)
\end{equation}
with $i = \bf {A, B, C, D}$, $H_i$ is the Hamiltonian at $i$ and $\rho_i$ is the corresponding density matrix. The sign convention followed in this paper is as follows: the amount of energy absorbed by the WM ($\mathcal{Q}_{in}$) is taken to be positive whereas energy released by the WM ($\mathcal{Q}_{out}$) is taken to be negative. 
%We denote the energy absorbed by the WM during the non-unitary stroke \textbf{A} $\rightarrow$ \textbf{B} with $\mathcal{Q}_{in}$  and the energy released by the WM during \textbf{C} $\rightarrow$ \textbf{D} non-unitary stroke with $\mathcal{Q}_{out}$. 
These energies can be calculated as follows :
\begin{eqnarray}
\mathcal{Q}_{in} &=& \mathcal{E}_B - \mathcal{E}_A \\
\mathcal{Q}_{out} &=& \mathcal{E}_D - \mathcal{E}_C.
\end{eqnarray}
The output work is $\mathcal{W} = -(\mathcal{Q}_{in} + \mathcal{Q}_{out})$.
		The quantum Otto cycle works as an engine if $\mathcal{Q}_{in} > 0, \mathcal{Q}_{out} < 0, \mathcal{W} < 0$ { which can be achieved by choosing $\lambda_1 > \lambda_2$ and
		$T_H>T_C$}. 
		The quantities of interest which characterize its performance are 
efficiency and the power output defined as
\begin{eqnarray}
\eta &=& -\frac{\mathcal{W}}{\mathcal{Q}_{in}}\\
\mathcal{P} &=& \frac{\mathcal{W}}{\tau_{\rm{total}}},
\end{eqnarray}
where $\tau_{\rm{total}} = \tau_1 + \tau_2 + \tau_H + \tau_C$.

Let us now outline {the main idea behind the technique and the }modification proposed during the unitary stroke \textbf{D} $\rightarrow$ \textbf{A}, which will help 
		in improving the performance of the engine. {If $T_c$ is set to zero  (or close to zero), the ideal state at $A$ should be the ground state (or very close to the ground state), provided the evolution from \textbf{D} to \textbf{A} is adiabatic. A non-adiabatic evolution results in a higher energy state at \textbf{A} which will reduce the performance of the engine.  
		We set the parameters of the Hamiltonian such that the low energy state at A is determined by $H_1$, which in most of the cases can be chosen to be a trivial state. In such a 
		situation, we can identify a Hamiltonian using which the state of the system may be evolved to the required state.
	 Thus, an additional evolution with this new Hamiltonian after the normal unitary evolution may help in achieving 
		higher efficiency along with higher power. We give the detailed steps below for a Hamiltonian where $H_0$ is able to bring in the required change.}

%We choose $\lambda_1 >> 1$ and $\lambda_2$ is such that $\lambda_2 < \lambda_c < \lambda_1$ with $\lambda_c$ being the critical point (CP). Therefore during the adiabatic strokes, the WM is driven such that it can cross the CP or not cross the CP. The crossing the CP generates non-adiabatic excitations which reduce the work output and efficiency of the engine. To overcome this problem, we search for ways through which this reduction in the performance can be avoided. We propose one way of achieving this by modifying one of the adiabatic stroke as described in the following section.
% 
%\section{Free evolution stroke}\label{secIII}
We modify the unitary stroke \textbf{D $\rightarrow$ 
A} by including the additional time evolution for a particular time after the usual unitary evolution, see figure \ref{fig_kickedcycle}. 
%		As discussed, we shall show that this modification will improve both, the efficiency as well as the power of the engine. 
		The exact protocol that we adopt is as follows :

\begin{figure}[h]
\centering
\includegraphics[clip, trim= 2cm 10cm 1cm 4cm,scale=.4]{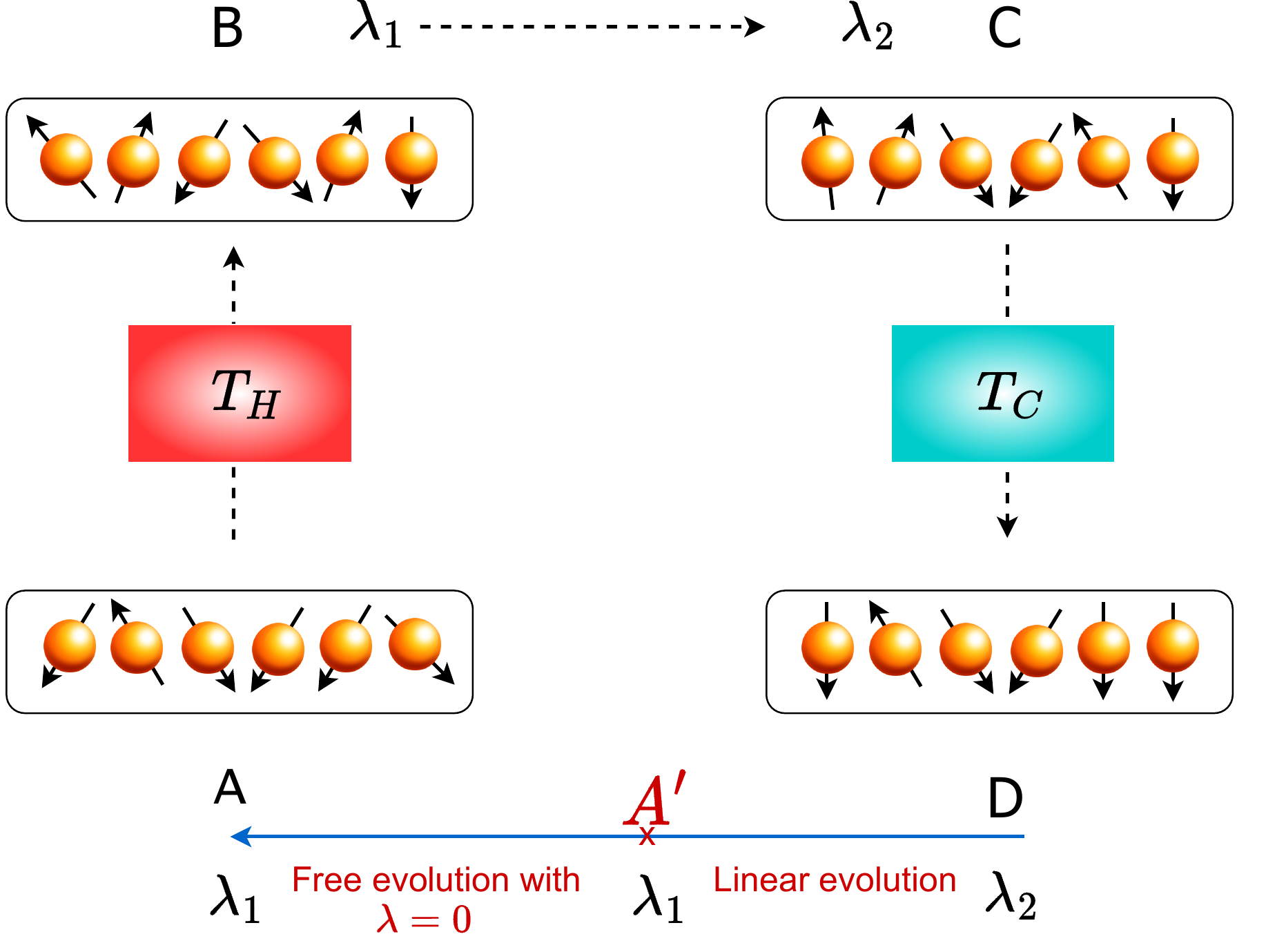}
\caption{{\bf Schematic diagram of the modified quantum Otto cycle.} At {\bf D} the system is evolved to ${\mathbf{A^{\prime}}}$ with a linear time dependence. At ${\mathbf{A^{\prime}}}$, $\lambda_1$ is switched off and the system freely evolves with $H_0$ for a time $\tau_k$ to reach {\bf A} where $\lambda_1$ is switched on.}
\label{fig_kickedcycle}
\end{figure} 

At {\bf D} the WM is decoupled from the cold bath and 
$\lambda_2$ is changed back to $\lambda_1$ with a time dependence
proportional to $t/\tau_2$ such that it reaches 
${\mathbf{A^{\prime}}}$. 
At ${\mathbf{A^{\prime}}}$, $\lambda_1$ is switched off and the 
system is allowed to evolve with $H_0$ for a time $\tau_k$. Henceforth, we shall call this evolution as free evolution. The parameter $\lambda$ is again switched back to $\lambda_1$ at {\bf A}
 after time $\tau_k$.  We shall refer to engines which follow such a modified Otto cycle as freely evolved engines. The genesis of this idea comes from the fact that the unitary evolution from {\bf D} to ${\mathbf{A^{\prime}}}$ will result in a state that is excited at ${\mathbf{A^{\prime}}}$ due to the non-adiabatic dynamics. 
% The idea of this modification is that
%a free evolution with $H_0$ should take the system to a state which has
% lower energy as compared to the one at ${\mathbf{A^{\prime}}}$. 
 Thus, the success
of this modification strongly depends on whether free evolution with $H_0$ is able to 
take the system to a lower energy state at ${\mathbf A}$. {In short, our protocol will work successfully only if $\mathcal{E}_A < \mathcal{E}_{A'}$, or 
$\langle H_0+ \lambda_1 H_1 \rangle _A < \langle H_0+ \lambda_1 H_1 \rangle_{A'}$, meaning the state at $A$ is closer to the ground state as compared to the state at $A'$}.
As we shall show below using examples, this is indeed possible. The WM is then connected  to the hot bath in the non-unitary stroke
 \textbf{A $\rightarrow$ B} after which the other strokes are 
followed.
\end{itemize} 

In the following sections we implement these ideas into the quantum Otto cycle using an integrable as well as a non-integrable WM.

\section{Integrable Model as WM}
\label{secIV}
Let us first consider the widely studied integrable model, the transverse field Ising model as the working medium. The Hamiltonian is given by
\begin{equation}\label{eqn_IsingHam}
H = -J \sum_n \sigma_{n}^{z} \sigma_{n+1}^z - h(t) \sum_n \sigma_{n}^x
\end{equation}
with $H_0 = -J \sum_n \sigma_{n}^{z} \sigma_{n+1}^z$ and $H_1 = -\sum_n \sigma_{n}^x$. Here, $\sigma_{n}^{\mu}$ with $\mu = x, y, z$ are the Pauli matrices at site $n$, $J$ is the nearest neighbour interaction strength and
$h(t)$ is the transverse field which corresponds to the parameter $\lambda$ for TIM. The model shows quantum phase transition from the paramagnetic state ($h \gg J$) to the ferromagnetic state ($h \ll J$) with the quantum critical point occurring at $h = \pm J$ \cite{lieb61two, pfeuty70the, bunder99effect}. The relaxation time diverges at the critical point. As a result, there will be excitations generated no matter how slow the parameter $h$ is varied, if the unitary strokes involve crossing the critical point \cite{sachdev_2011, dutta15quantum, RevModPhys.83.863}. 

\begin{figure}[h]
\begin{subfigure}{.5\textwidth}
  \centering
  % include first image
  \includegraphics[width=.95\linewidth]{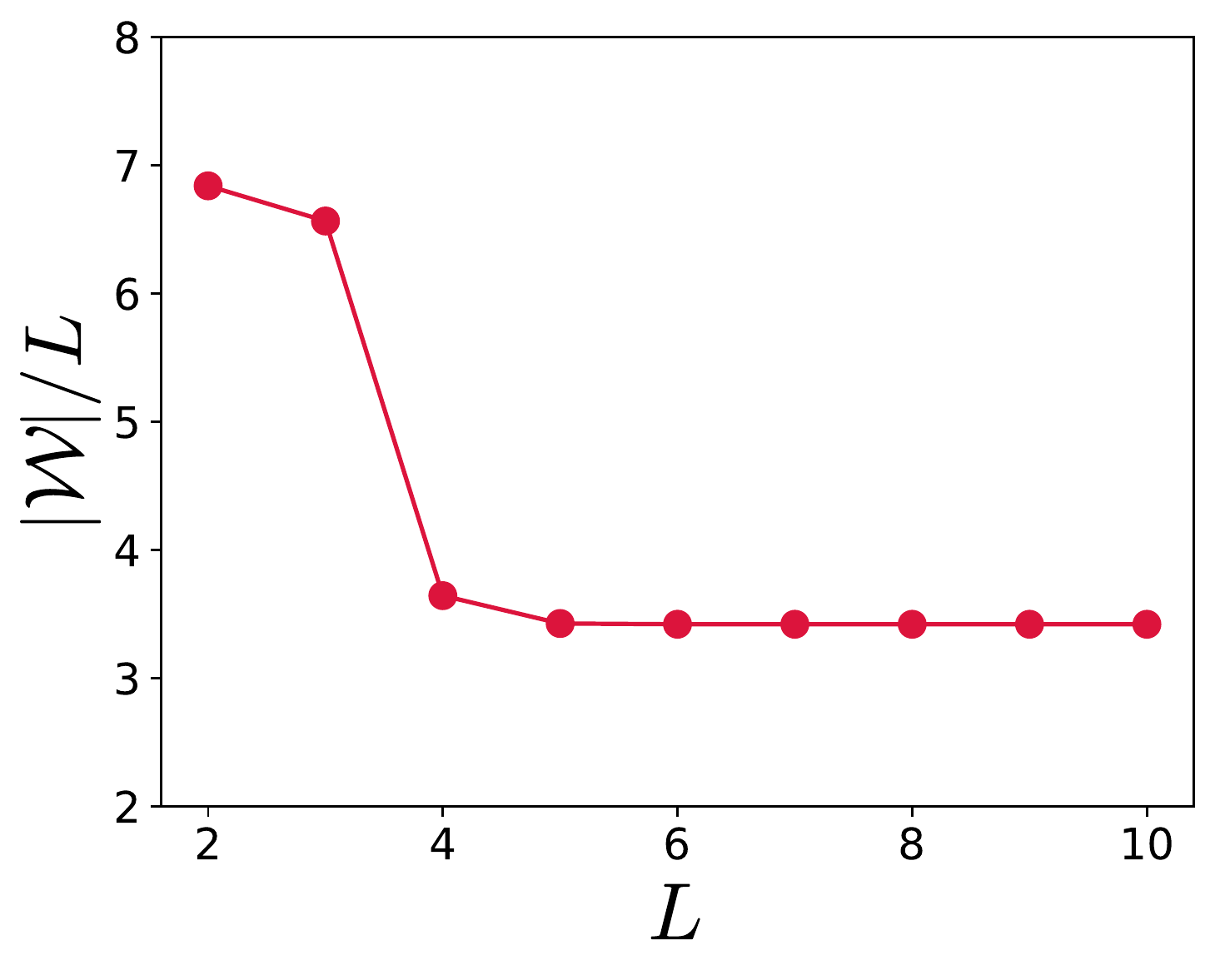}  
  \caption{}
  \label{fig_tim_wVtauka}
\end{subfigure}
\begin{subfigure}{.5\textwidth}
  \centering
  % include first image
  \includegraphics[width=.95\linewidth]{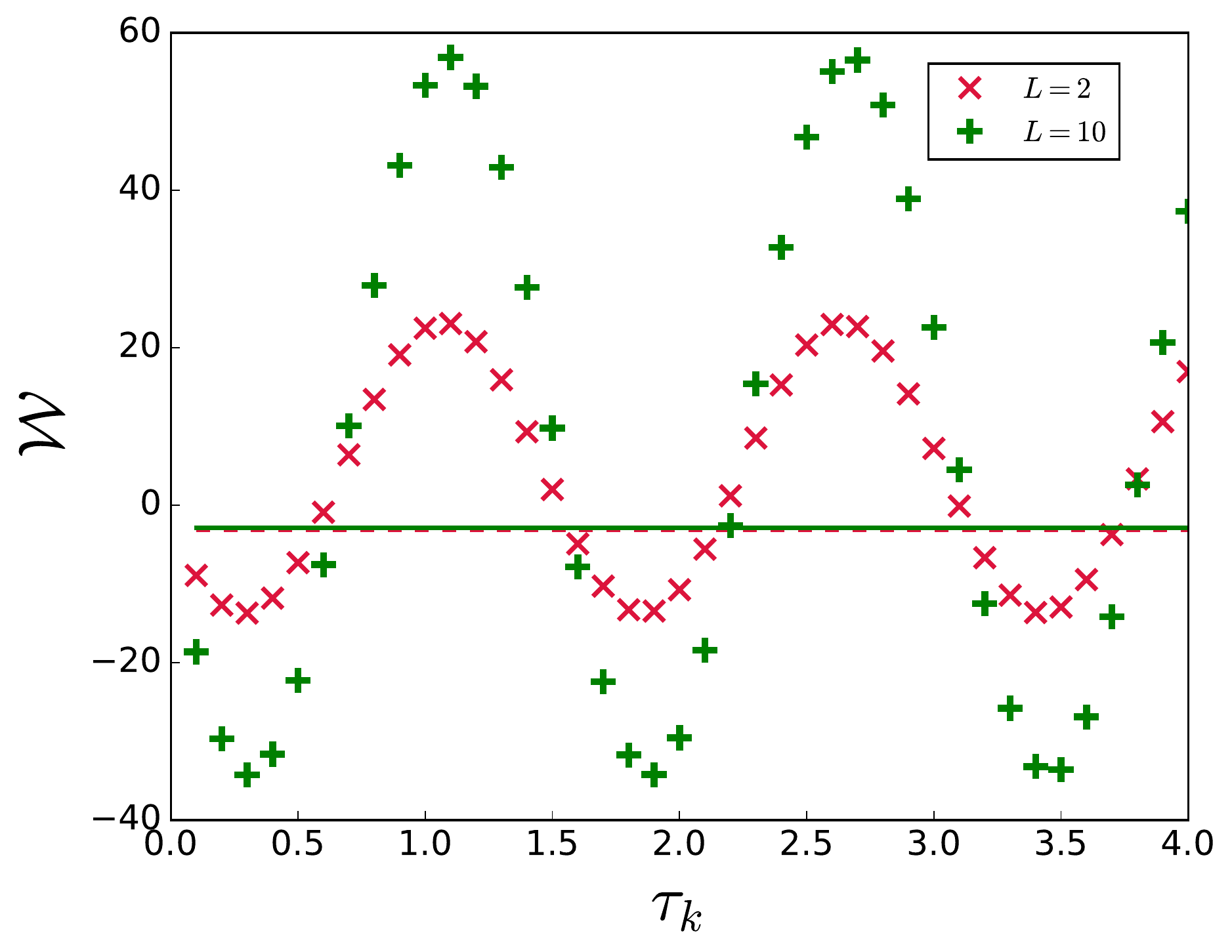}  
  \caption{}
  \label{fig_tim_wVtaukb}
\end{subfigure}
%\centering
%\includegraphics[scale=.4]{Fig_2n.pdf}
\caption{(a) {\bf $\mathcal{|W|}/L$ as a function of $L$ showing that the work output is proportional to the system size for large system sizes}. (b) {\bf $\mathcal{W}$ as a function of $\tau_k$ for different system sizes for TIM.} The data points correspond to freely evolved engines and the dashed line and the continuous line correspond to normal finite time engines of $L = 2$ and $L = 10$ respectively. Note that the dashed and continuous lines are constant since there is no $\tau_k$ dependence for the normal engines. The parameters are $h_1 = 10, h_2 = 0.2, T_H = 100, T_C = 0.001, \tau_1 = 0.1, \tau_2 = 0.1.$ The $\tau_k$ at which the maximum work is done by the engine, is also seen to be system size independent. }
\label{fig_tim_wVtauk}
\end{figure}

We now need to select a protocol to vary $h$ from $h_1$ to $h_2$ during the unitary strokes. The driving protocol for the {\bf B} to {\bf C} stroke is chosen to be
\begin{equation}\label{eqn_protocol_bc}
h(t) = h_1 + (h_2 - h_1)\frac{t}{\tau_1}, \hspace{0.5cm} 0 < t < \tau_1.
\end{equation}
We have taken $h_1 \gg J$ and $h_2 < h_1$ so that the cycle works as an engine.
In the case of the freely evolved engine, once the system reaches the thermal state at {\bf D} with $h=h_2$ and $T = T_C$, it is decoupled from the cold bath after which the transverse field is changed from $h_2$ to $h_1$ to reach ${\mathbf{A^{\prime}}}$ following
\begin{equation}\label{eqn_protocol_da}
h(t) = h_2 + (h_1 - h_2)\frac{(t - a)}{\tau_2}, \hspace{0.5cm} a < t < a + \tau_2, 
\end{equation}
where $a =  \tau_C + \tau_1$. At ${\mathbf{A^{\prime}}}$, the transverse field is switched off and the WM is allowed to evolve freely with $H_0$ till a time $\tau_k$. The transverse field is switched back to $h_1$ at $\tau_k$ which corresponds to point \textbf{A} in the Otto cycle after which it is connected to the hot bath at $T_H$, and the cycle repeats.
For numerical calculations, we set $J =1$ and use periodic boundary conditions. We first plot the output work of the engine per system size as a function of system size in figure (\ref{fig_tim_wVtauka}). Clearly, $\mathcal{|W|}/L$ goes to a constant from system size $L = 5$ onwards indicating that the work output is proportional to the system size for large system sizes. $\mathcal{W}$ is plotted as a function of $\tau_k$ for $L = 2$ and $L=10$ in figure (\ref{fig_tim_wVtaukb}). It is evident from this figure that for specific ranges of $\tau_k$ values, the output work of the freely evolved engine reaches a more negative value compared to that of normal finite time engines. Clearly, the magnitude of the work output by an engine is maximum when $\mathcal{E}_A$ is the minimum and that happens for optimal values of $\tau_k$. Identifying the optimal value of $\tau_k$ is important since it decides how much the work output can be increased which is same as how much the energy at \textbf{A} can be reduced. To get a better understanding of this optimal $\tau_k$, below we calculate $\tau_k$ for a two spin system analytically in the adiabatic limit.

\subsection{Two spin case}
For a two spin system, the value of $\tau_k$ can be derived analytically in the adiabatic limit. The Hamiltonian for a two spin system at {\bf D} in the $z$ basis 
with periodic boundary conditions is given by
\begin{equation}
H_D = \begin{bmatrix}
 -2J & -h_2 & -h_2 & 0 \\ -h_2 & 2J & 0 & -h_2\\ -h_2 & 0 & 2J & -h_2\\ 0 & -h_2 & -h_2 & -2J
\end{bmatrix},
\end{equation}
with eigenenergies $\epsilon_1, \epsilon_2, \epsilon_3, \epsilon_4$.
Since the WM is in the thermal state corresponding to $T_C$ at {\bf D}, the density matrix in the eigenbasis can be written as
\begin{equation}
\rho_D = \begin{bmatrix}
 \frac{e^{-\beta_C \epsilon_1}}{Z} & 0 & 0 & 0 \\ 0 & \frac{e^{-\beta_C \epsilon_2}}{Z} & 0 & 0\\ 0 & 0 & \frac{e^{-\beta_C \epsilon_3}}{Z} & 0\\ 0 & 0 & 0 & \frac{e^{- \beta_C \epsilon_4}}{Z}
\end{bmatrix},
\end{equation}
where $\beta_C = \frac{1}{ T_C}$ with $k_B  = 1$.
Considering adiabatic evolution from \textbf{D $\rightarrow$} ${\mathbf{A^{\prime}}}$ which corresponds to a large $\tau_2$, we can write $\rho_D = \rho_{A^{\prime}}$ in their respective eigenbasis. The energy at ${\mathbf{A^{\prime}}}$ can be calculated as
\begin{eqnarray}
\mathcal{E}_{A^{\prime}} &=& \text{Tr} (H_A{^{\prime}} \rho_A{^{\prime}}) \\
&=& -4 h_1 \alpha - \frac{4J^2 \alpha}{h_1} + 4J \delta
\end{eqnarray}
with 
\begin{eqnarray}
\alpha &=& \frac{h_1 \sinh \left(\frac{2\sqrt{h_2^2 + J^2}}{T_C} \right)}{2\sqrt{h_1^2 + J^2} \left( \cosh\frac{2J}{T_C} + \cosh\frac{2\sqrt{h_2^2 + J^2}}{T_C}\right)},\\
\delta &=& \frac{-\sinh(\frac{2J}{T_C})}{2 \left( \cosh\frac{2J}{T_C} + \cosh\frac{2\sqrt{h_2^2 + J^2}}{T_C}\right)}.
\end{eqnarray}
We now evolve $\rho_A{^{\prime}}$ with $U = \exp(-i (-J  \sigma_i^z \sigma_{i+1}^z)t)$ up to $\tau_k$ to reach \textbf{A} so that the energy at {\bf A} takes the form
\begin{align}\label{eqn_ee_analytic}
\mathcal{E}_A = -4 h_1 \alpha \cos(4J \tau_k) - \frac{4J^2 \alpha}{h_1} \cos(4 h_1) \nonumber\\
+ 4J \delta -4 J \alpha \sin(4h_1) \sin(4J \tau_k),
\end{align}
%\end{equation}
where $H_{A} = H_{A^{\prime}}$.

\begin{figure}[h]
\centering
\includegraphics[scale=.4]{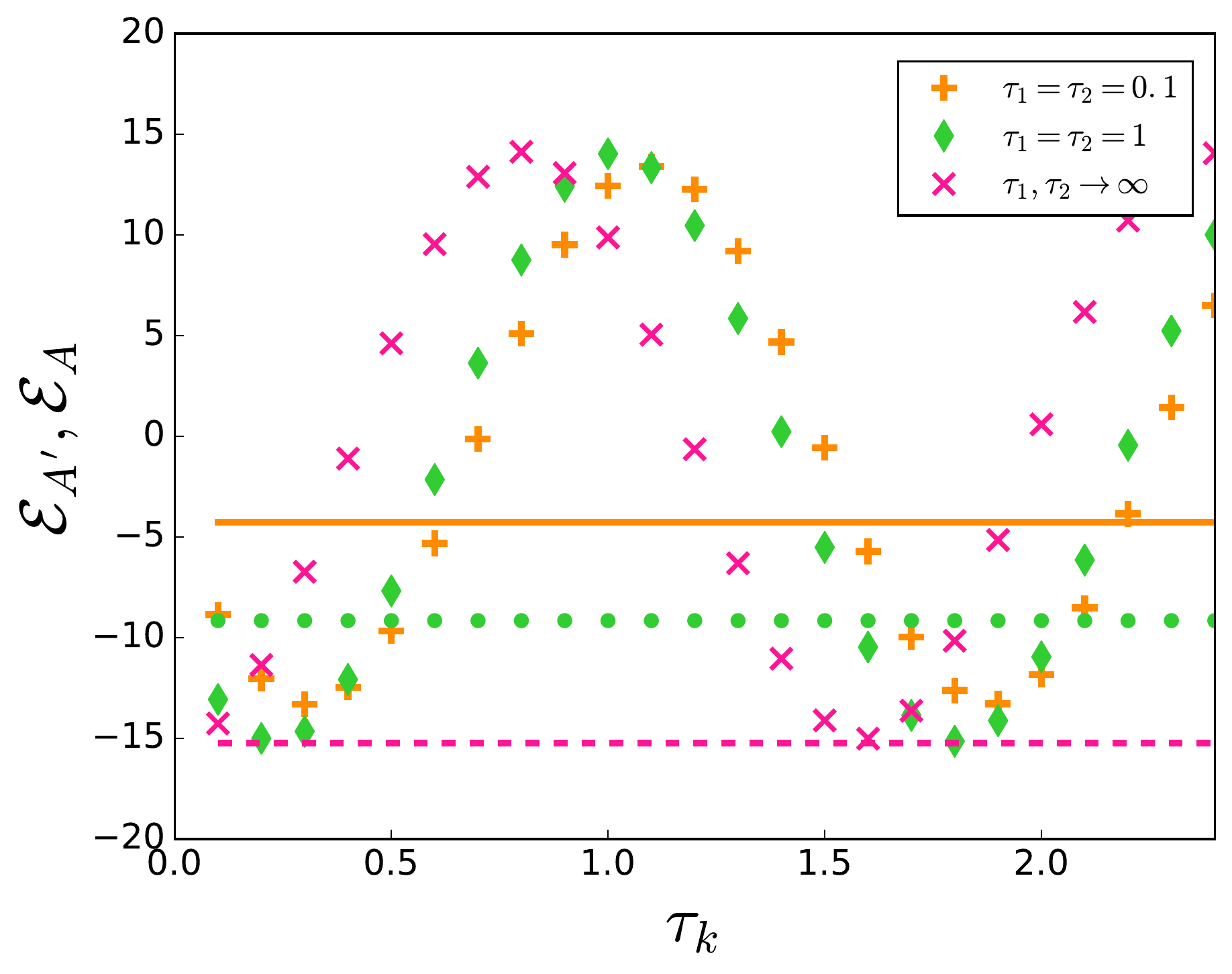}
\caption{{\bf $\mathcal{E}_{A}$ as a function of $\tau_k$ for $L=2$ comparing adiabatic engine with finite time engines.} Here, the data points correspond to $\mathcal{E}_A$ for different $\tau_2$ as described in the legend whereas the continuous, dotted and dashed line correspond to $\mathcal{E}_{A^{\prime}}$ for $\tau_1 =\tau_2 = 0.1$, $\tau_1,\tau_2 = 1$ and $\tau_1,\tau_2 \rightarrow \infty$, respectively. The parameters are $L = 2, h_1 = 10, h_2 = 0.1, T_H = 100, T_C = 0.01$. }
\label{fig_tim_analytic}
\end{figure}

Since we aim to obtain the $\tau_k$ for which $\mathcal{E}_{A}$ is minimum, we optimize  $\mathcal{E}_{A}$ with respect to $\tau_k$ to get
\begin{equation}\label{eqn_tau_opt}
\tan(4J\tau_k) = \frac{J}{h_1} \sin(4 h_1)
\end{equation}
so that minimum $\mathcal{E}_{A}$ occurs at $\tau_k^{opt} =  n \pi/4$ with $n = 0, 2, 4,..$ for $h_1 \gg 1$ and $J = 1$. We plot $\mathcal{E}_{A}$ obtained in equation (\ref{eqn_ee_analytic}) with $\tau_k$ and compare with finite time engines in figure (\ref{fig_tim_analytic}). It is clear that for an adiabatic evolution from \textbf{D} to ${\mathbf{A^{\prime}}}$, i.e., $\tau_2 \rightarrow \infty$, the $\tau_k^{opt}$ where the minimum $\mathcal{E}_A$ occurs is at $\tau_k = 0$. It is to be noted that the adiabatic evolution is the best possible way of evolution by which no additional excitations occur in the system. Therefore, any further free evolution will only take the system to a higher energy state which is also seen in figure (\ref{fig_tim_analytic}). On the other hand, for a finite time evolution from \textbf{D} to ${\mathbf{A^{\prime}}}$, the state reached at ${\mathbf{A^{\prime}}}$ need not be the state with the lowest possible energy as obtained through adiabatic evolution. Hence, there is a possibility that free evolution from ${\mathbf{A^{\prime}}}$ to \textbf{A} can take the system to a lower energy state. From figure (\ref{fig_tim_analytic}), for finite time evolution with $\tau_2 = 0.1$, the $\tau_k^{opt}$ value occurs at 0.3. Also, we observe that when $\tau_2$ increases to 1, the $\tau_k$ value at which the minima of $\mathcal{E}_A$ occurs shifts closer and closer to the adiabatic minima of $\tau_k^{opt}=0$.

The optimal $\tau_k$ value obtained for the two spin case is the same for larger system sizes as also seen in figure (\ref{fig_tim_wVtaukb}). In \ref{appA}, we analytically calculate the value of optimal $\tau_k$ for different system sizes in the adiabatic limit and show that this $\tau_k^{opt}$ is indeed independent of system size. {Therefore, the knowledge of $\tau_k^{opt}$ from a two spin system is sufficient for performing the free evolution technique in higher spin systems in the example considered.}

\begin{figure}[h]
\centering
\includegraphics[scale=.4]{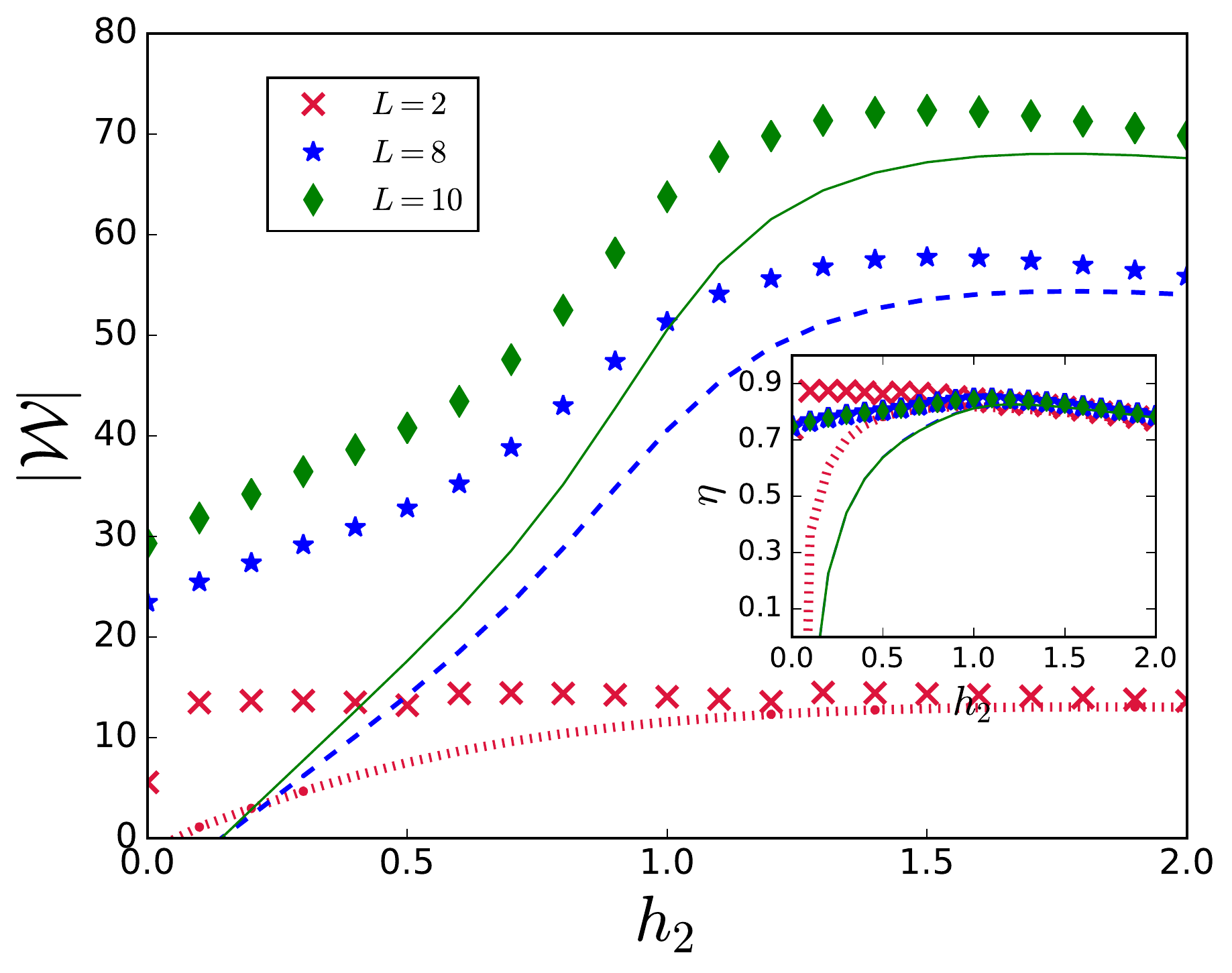}
\caption{{\bf $\mathcal{|W|}$ as a function of $h_2$ for different system sizes of TIM.} The data points correspond to freely evolved engines for different system sizes as shown in the legend, and the dotted, dashed and continuous line correspond to the work done by the finite time engines of system sizes $L=2$, $L=8$ and $L=10$, respectively. Inset: $\eta$ as a function of $h_2$ 
with the same legend as in the main figure. The parameters are $h_1 = 10, T_H = 100, T_C = 0.001, \tau_1 = 0.1, \tau_2 = 0.1.$ }
\label{fig_tim_workVsh2}
\end{figure}

We now study the variation of output work and efficiency for the freely evolved engine with $\tau_k$ set to $\tau_k^{opt}$ and compare it with usual finite time engines with same $\tau_1$ and $\tau_2$. 
Figure (\ref{fig_tim_workVsh2}) shows the output work as a function of $h_2$ for different system sizes. In the engines where $h_2$ is close to the critical point or smaller so that the critical point is crossed, 
we see that the additional free evolution of the engine improves the work output of the engine to a great extent for all system sizes. Similar behavior is also seen in the efficiency plot (inset of figure (\ref{fig_tim_workVsh2})). This is due to the fact that the  energies at ${\mathbf{A^{\prime}}}$ and {\bf A} follow the inequality $\mathcal{E}_A < \mathcal{E}_{A^{\prime}}$ resulting to an increase in the magnitude of $\mathcal{Q}_{in}$ which further reflects in the improved output work and efficiency.

\begin{figure}[h]
\centering
\includegraphics[scale=.4]{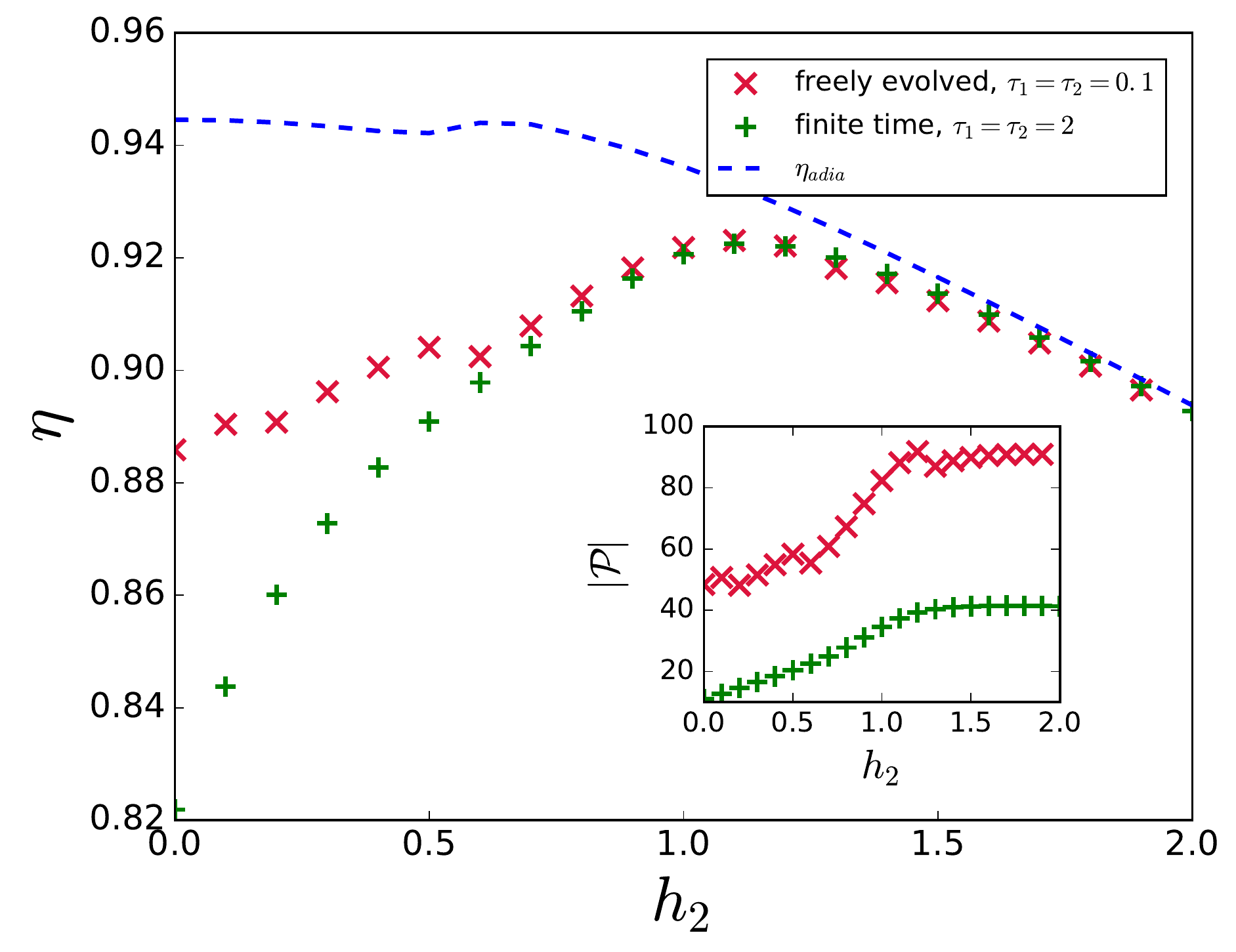}
\caption{{\bf $\eta$ as a function of $h_2$ for $L= 10$ TIM.} Inset: $\mathcal{|P|}$ as a function of $h_2$. The parameters are $h_1 = 20, T_H = 1000, T_C = 0.001$. We have set $\tau_H + \tau_C = 0.2$. The values of $\tau_1$ and $\tau_2$ for the freely evolved engine is intentionally chosen to be lesser than that of normal finite time engine to highlight the advantage of free evolution. For the freely evolved engine, optimal $\tau_k$ value is $h_2$ dependent. }
\label{fig_tim_PVsh2}
\end{figure}

Adiabatic evolution ($\tau_1,\tau_2 \rightarrow \infty$) leads to highly efficient engines but at the cost of output power. As can be easily verified, increasing $\tau_1$ and $\tau_2$ enables finite time engines to reach adiabatic efficiency value $\eta_{adia}$. But this clearly reduces the output power due to large values of $\tau_1$ and $\tau_2$. In contrast, the freely evolved engines have high efficiency closer to $\eta_{adia}$ even at small values of $\tau_1$ and $\tau_2$ (with free evolution time $\tau_k^{opt}$) along with finite power. {Note that the expression for the total cycle time gets modified to
$\tau_1+\tau_2+\tau_C+\tau_H+\tau_k^{opt}$. But we can still get better power and efficiency by taking smaller values of $\tau_1$ and $\tau_2$ so that the work done is improved due to non zero
$\tau_k^{opt}$}. This can be seen from figure (\ref{fig_tim_PVsh2}) where we show that the efficiency as well as the power improves in case of the freely evolved engines with smaller $\tau_1, \tau_2$ as compared to normal finite time engines with larger $\tau_1$ and $\tau_2$. Here, $\tau_H$ and $\tau_C$ can be controlled by changing the coupling strength of the system with the bath. Therefore, we have taken comparable values of 
thermalization times as well as times for the unitary strokes so that the power output 
of the engine can be maximized.
%In other words, we get better efficiency and output power with smaller total cycle time ($\tau_1 + \tau_2 +  \tau_H + \tau_C +\tau_k$) compared to usual finite time engines where decreasing cycle time will lead to increase in output power but with reduced efficiency.

We also extend the technique of free evolution to the momentum ($k$) space which allows us to go to higher system sizes. The Hamiltonian in equation (\ref{eqn_IsingHam}) when written in $k$ space takes the form \cite{revathybatheng}
\begin{equation}
H = \sum_{k>0} \psi_k^{\dagger} H_k \psi_k
\end{equation} 
with $\psi_k^{\dagger} = (c_k^{\dagger}, c_{-k})$ and 
\begin{equation}\label{Ham_new}
H_{k} = \begin{bmatrix}
 2(h(t) + \cos k) & 0 & 0 & 2 \sin k \\ 0 & 0 & 0 & 0\\ 0 & 0 & 0 & 0\\ 2 \sin k & 0 & 0 & -2(h(t) + \cos k) 
\end{bmatrix}.
\end{equation}
The unitary dynamics that is undergone in the \textbf{B} to \textbf{C} and \textbf{D} to \textbf{A} strokes is described by the von-Neumann equation and the density matrix at {\bf B} and {\bf D} for each $k$ mode will be the thermal state corresponding to $T_H$ and $T_C$ respectively, as given in Ref. \cite{revathybatheng}. For the free evolution, the density matrix at \textbf{A} for each mode can be written as
\begin{equation}
\rho_k^A = U_k \rho_k^{A^{\prime}} U_k^{\dagger},
\end{equation}
with $U_k = \exp(-i H^{\prime}_k \tau_k)$ and $H^{\prime}_k = H_k$ with $h(t)=0$.  
%As discussed before, the $\tau_k^{opt}$ is independent of system size which has been verified numerically in momentum space. This further helps in extending the technique to larger system sizes. 
\begin{figure}[h]
\centering
\includegraphics[scale=.4]{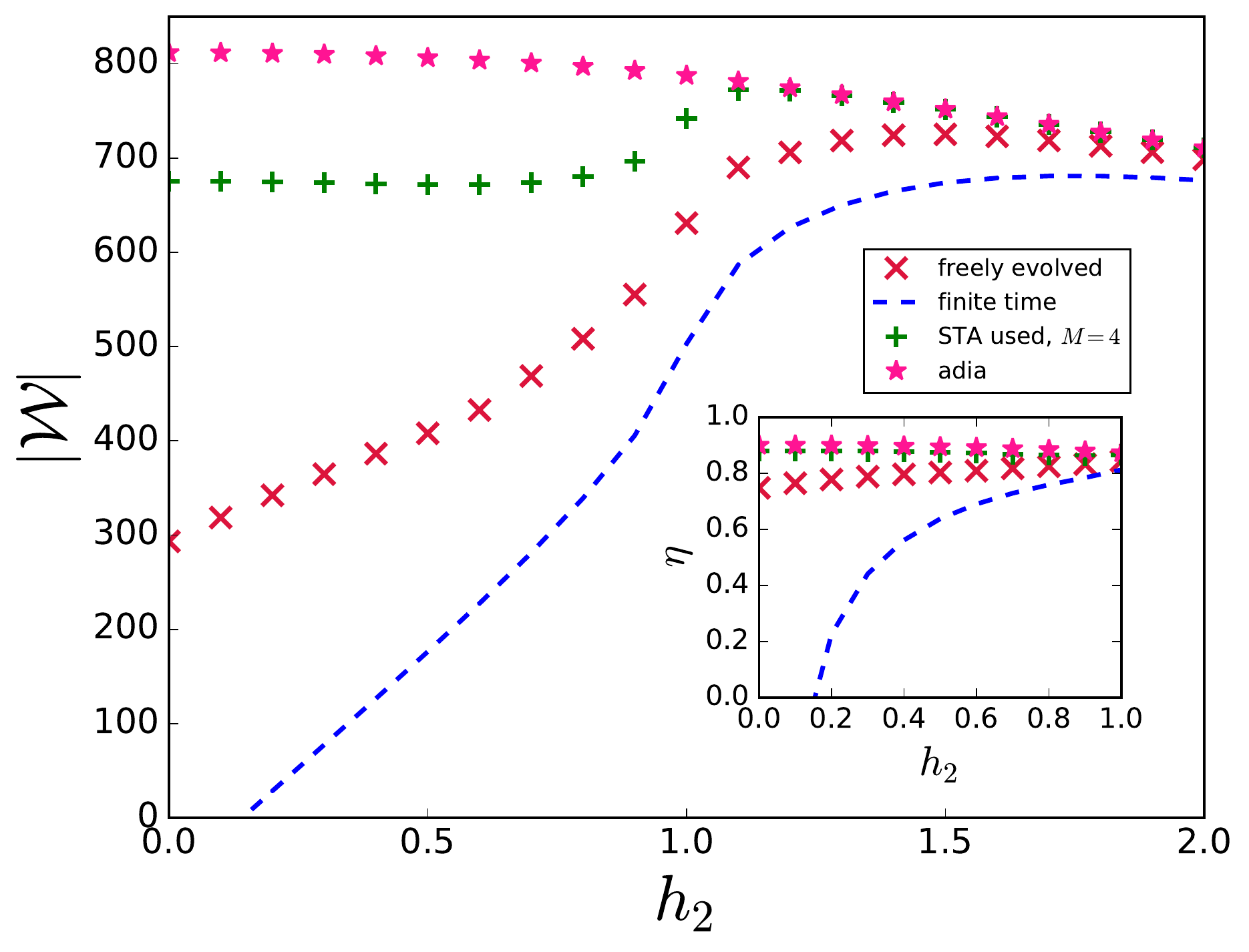}
\caption{{\bf $\mathcal{|W|}$ as a function of $h_2$ for $L= 100$ TIM.}  The parameters are $h_1 = 10, T_H = 100, T_C = 0.001, \tau_1 = 0.1, \tau_2 = 0.1$. We have set $\tau_H + \tau_C = 0.2$. Inset: $\eta$ as a function of $h_2$ with the same legend as in the main figure. For the freely evolved engine, optimal $\tau_k$ value is $h_2$ dependent. The control Hamiltonian for STA is truncated to $M$ body terms with $M=4$ \cite{PhysRevLett.109.115703}.}
\label{fig_tim_L100Vsh2}
\end{figure} 
In figure \ref{fig_tim_L100Vsh2}, we plot the work output and efficiency (in the inset) by the freely evolved engines as a function of $h_2$ for $L = 100$ and compare it with normal finite time engine, finite time engine using STA, and adiabatic engine. As expected, the freely evolved engine performs better than the finite time engine without any control. On the contrary, its performance is lower when compared with finite time engines using STA. But we emphasize here the simplicity of implementing this protocol as compared to STA which makes the free evolution technique practical.

We further extend the technique of free evolution to a non-integrable model in the next section and show the effectiveness of this protocol.

\section{Non integrable Model as WM}
\label{secV}
In this section we use a non-integrable model, the antiferromagnetic transverse Ising model with longitudinal field (LTIM) to demonstrate the freely evolved engine. The Hamiltonian is given by \cite{PhysRevB.92.104306, PhysRevE.99.012122}
\begin{equation}
H = J \sum_n \sigma_{n}^{z} \sigma_{n+1}^z - h(t) \sum_n \sigma_{n}^x - B_z \sum_n \sigma_{n}^z ,
\end{equation}
where $J$ is the antiferromagnetic interaction , $B_z$ is the longitudinal field and $h$ is the time dependent transverse field which is varied as given in equations (\ref{eqn_protocol_bc}) and (\ref{eqn_protocol_da}) during the unitary strokes. 

In the case of LTIM, $H_0 = J \sum_n \sigma_{n}^{z} \sigma_{n+1}^z -  B_z \sum_n \sigma_{n}^z$ and $H_1 = - h(t) \sum_n  \sigma_{n}^x$ so that $h$ is switched off to zero between ${\mathbf{A^{\prime}}}$ and {\bf A} for a time $\tau_k$. At {\bf A}, $h(t)$ is again switched back to $h_1$ and the cycle repeats.

\begin{figure}[h]
\centering
\includegraphics[scale=.4]{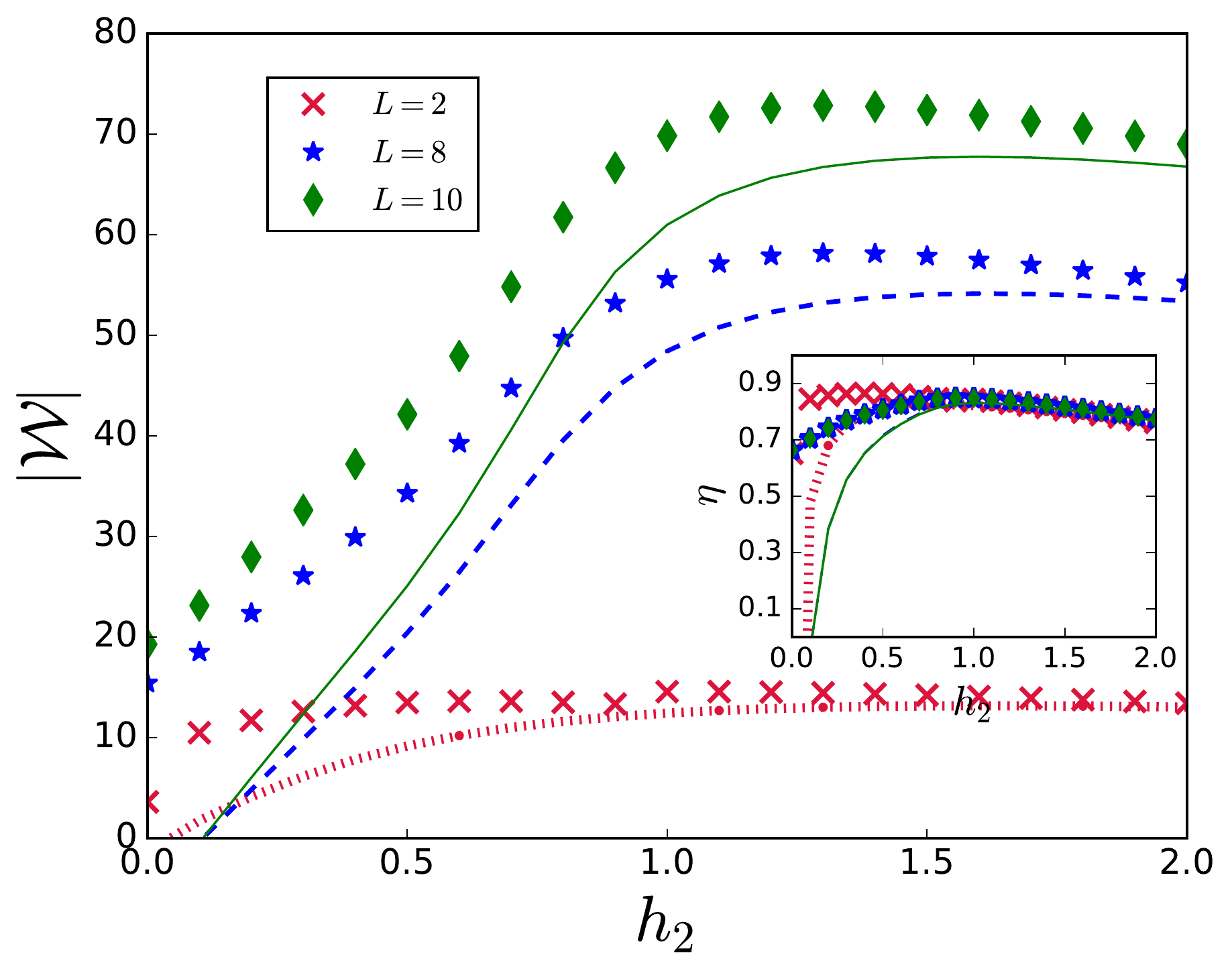}
\caption{{\bf $\mathcal{|W|}$ as a function of $h_2$ for different system sizes in case of LTIM.} The data points correspond to freely evolved engines for different system sizes as shown in the legend, and the dotted, dashed and continuous line correspond to the work done by the finite time engines of
system sizes $L=2$, $L=8$ and $L=10$, respectively. The parameters are $B_z = 1, h_1 = 10, T_H = 100, T_C = 0.001, \tau_1 = 0.1, \tau_2 = 0.1$. Inset: $\eta$ as function of $h_2$ with the same legend as in the main figure.  }
\label{fig_ltim_workVh2}
\end{figure}

In figure (\ref{fig_ltim_workVh2}), the work output of the freely evolved engine as well as normal finite time engines is plotted as a function of $h_2$ for different system sizes of LTIM. Similar to the case of TIM, we observe a significant improvement in the work output of the freely evolved engine, especially for small values of $h_2$ where the energy gaps are small. As mentioned before, here also $\tau_k^{opt}$ is different for different $h_2$. 
The improvement in the performance of the engine is also visible in terms of its efficiency, as shown in the inset of figure (\ref{fig_ltim_workVh2}).

%the output work and efficiency (inset) of the free evolved engine is compared with normal finite time engines. In the case of free evolved engine, we observe a significant improvement in the performance of the engine for small values of $h_2$ where the energy gaps are small. As mentioned before, here also $\tau_k^{opt}$ is different for different $h_2$.

The $\tau_k^{opt}$ value for different system sizes can be found out analytically in the adiabatic limit and is presented in \ref{appB}. For the case of LTIM also, the $\tau_k^{opt}$ is found to be system size independent.

\section{Conclusions}
\label{secVI}
Through this work, we propose a simple and effective way to improve the work output and efficiency of a quantum Otto engine which uses a quantum substance as the working medium. The excitations produced 
during the operation of the engine {due to crossing the quantum critical point of the WM adversely affects}  its performance. {These excitations can be reduced using the 
simple method that we propose. In the free evolution example considered in this paper,
we modify one of the unitary strokes by first letting it undergo the usual unitary dynamics of the Otto cycle followed by an evolution 
with only one of the terms of the Hamiltonian for a time $\tau_k$. This is achieved  by switching off the other term in the Hamiltonian for a time $\tau_k$ (carefully chosen as discussed below) 
after which it is switched back to its original value.}
 Our results using the integrable as well as the non-integrable model shows that 
this method helps significantly in improving the performance of the engine. 

The main idea or the bigger picture here is to evolve the system between $A'$ and $A$ 
using some time independent $\tilde H$ which will reduce the excitations (or reduce the energy) at $A$ as compared to $A'$. In general, this $\tilde H$ can be part of the total $H$ as is the case in this work, 
or choose a different $\tilde H$ which can reduce the energy at $A$. One way to do this is to choose $\tilde H$ which does not commute with $H$ to introduce some non-trivial dynamics. In other words,
$\tilde H$ should be choosen such that it can increase the probability for the working medium to be found in the ground state. 
%The method of switching on and off of a term of the Hamiltonian without introducing 
%any new term is the simplest way to achieve this due to non-commutativity of the two terms in the $H$, as is also shown in Appendix \ref{appA}. 
The method of switching off and switching on of one of the terms 
of the Hamiltonian which do not commute with other terms in the 
Hamiltonian is the simplest way to achieve 
this. We present this particular aspect in this paper where
we show that such a change in the unitary stroke can increase 
the probability for the system to be found in the ground state of 
$H$, as also shown in \ref{appA}.

The next non-trivial point is for how long 
one should evolve the system between $A$ and $A'$, i.e., the value of $\tau_k$. We have calculated analytically the value of $\tau_k^{opt}$ for the  two systems under consideration and found it to be system 
size independent, which is also confirmed numerically. Unfortunately, we could not arrive at a general expression for $\tau_k$ in the case of a general $\tilde H$. We plan to explore this point in our future 
studies.

Unlike the conventional techniques such as shortcuts to adiabaticity or the  bath engineering techniques which 
may involve some additional time dependent terms for their implementation, our method just requires switching off of the time dependent parameter in the system Hamiltonian and evolving the system with a time
independent part of the system Hamiltonian for a specific time  making it easier to implement.
% without any additional costs.
% except for the energy costs that is associated with the mean energy change while switching off 
%and switching on $\lambda_1$. 
Experimental implementation of this protocol is also straightforward since it only involves tuning of the transverse field or some terms already present in the Hamiltonian. 
%This makes the technique easy to implement.

{In conclusion, our aim here is to highlight the possibility of improvement in the performance of an engine through a simple technique using certain examples. We would like to extend these 
results to the case of the most general Hamiltonian in our future studies.}

%This protocol of modification of the unitary stroke  involving free evolution of the system with some of the terms of the original Hamiltonian, has itself showed improvement in the performance of the engine to a large extent. However, we do acknowledge that this may not be always true. In such cases, the knowledge of the adiabatic state at \textbf{A} will be helpful in wisely choosing a new and different Hamiltonian for the free evolution between ${\mathbf{A^{\prime}}}$ and \textbf{A} such that it takes the system very close to its adiabatic state at \textbf{A}. 

\ack
We thank Victor Mukherjee for useful discussions and comments on the manuscript.

\section*{Data availability statement}
Any data that support the findings of this study are included within the article.

\appendix
\section{Finding $\tau_k^{opt}$ for different system sizes : TIM}
\label{appA}
We find an analytical estimate of the optimal $\tau_k$ value for different system sizes, in the adiabatic limit.
Consider the cold bath temperature to be zero so that the system reaches the ground state at {\bf D}. Let the system evolves adiabatically from \textbf{D $\rightarrow$ $\mathbf{{A^{\prime}}}$} which takes it to the ground state at ${\mathbf{A^{\prime}}}$. Since $h_1 \gg 1$, the state at {\bf ${A^{\prime}}$} can be written as
\begin{equation}
|\psi \rangle_{A^{\prime}} = \otimes_{L} |\rightarrow \rangle
\end{equation}
with all spins aligned along the $x$- direction.
The system is then allowed to evolve freely with $H_0$ with an evolution operator $U$ given by 
\begin{eqnarray}
U &=& \exp(-i (-J \sum_n \sigma_n^z \sigma_{n+1}^z) \tau_k)\\
&=& \displaystyle \prod_{n} \left( \cos(J \tau_k) + i\sin(J \tau_k) \sigma_n^z \sigma_{n+1}^z \right),
\end{eqnarray}
so that the state reached at {\bf A} for a two spin system is
\begin{equation}
|\psi \rangle_A = \cos(2J \tau_k) |\rightarrow \rightarrow \rangle + i \sin(2J\tau_k) |\leftarrow\leftarrow\rangle.
\end{equation}
Similarly, the state at {\bf A} for a 4-spin system is
\begin{eqnarray}
|\psi \rangle_A &=& \left(\cos^4(J \tau_k) + \sin^4(J\tau_k)\right) |\rightarrow\rightarrow\rightarrow\rightarrow\rangle \nonumber\\
&&- 2 \sin^{2}(J \tau)\cos^{2}(J \tau) |\leftarrow\leftarrow\leftarrow\leftarrow\rangle  \nonumber\\
&& -2 \sin^2 (J \tau) \cos^2(J \tau) (|\leftarrow\rightarrow\leftarrow\rightarrow\rangle + |\rightarrow\leftarrow\rightarrow\leftarrow\rangle) \nonumber\\
&& ( i \sin(J \tau)\cos^3(J \tau)- i \sin^3(J \tau)\cos(J \tau)) \nonumber\\
&& (|\rightarrow\rightarrow\leftarrow\leftarrow \rangle + |\leftarrow\leftarrow\rightarrow\rightarrow\rangle \nonumber\\
&&+ |\rightarrow\leftarrow\leftarrow\rightarrow\rangle + |\leftarrow\rightarrow\rightarrow\leftarrow\rangle).
\end{eqnarray}
Since the ground state is all spins along the $x$- direction, the probability for the system to be in its ground state at {\bf A} is given by $\cos^2(2J\tau_k)$ for $L=2$ and $(\cos^4(J \tau_k) + \sin^4(J\tau_k))^2$ for $L=4$. 
Our goal is to take the system to the lowest possible energy state at {\bf A}. Therefore we maximize the probability for the system to be in the ground state, to obtain the optimal $\tau_k$ at $\tau_k^{opt} = n \pi/4 J$ with $n = 0, 2, 4,..$ for both $L = 2$ and $L = 4$.  We have checked numerically and found it to be true for larger system sizes as well and thus the $\tau_k^{opt}$ is the same for all $L$. We have also verified that the probability for a system of size $L$ to be in its ground state can be calculated as $|a|^2$ with $ a = \cos(J \tau_k)^L + (i \sin(J \tau_k))^L$.

%The maximas for these two functions occur at the same $\tau_k$ and thus 

\section{Finding $\tau_k^{opt}$ for LTIM}
\label{appB}

\begin{figure}[h]
\centering
\includegraphics[scale=.4]{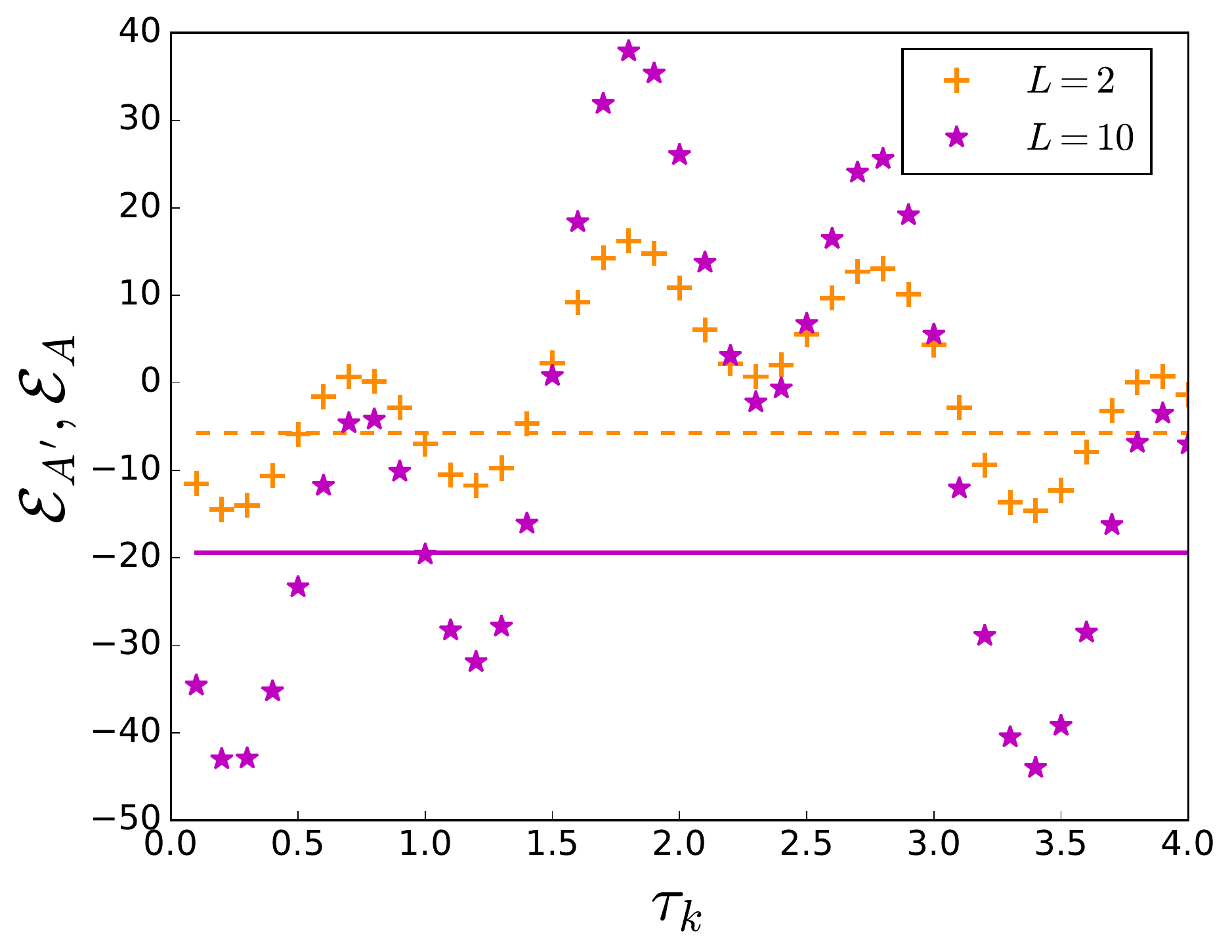}
\caption{{\bf $\mathcal{E}_{A}$ as a function of $\tau_k$ for different system sizes.} Here dashed line and continuous line represent $\mathcal{E}_{A^{\prime}}$ for $L =2$ and $L=10$, respectively. The parameters are $B_z = 1, h_1 = 10, h_2 = 0.1, T_H = 100, T_C = 0.001, \tau_1 = 0.1, \tau_2 = 0.1$.}
\label{fig_ltim_ee}
\end{figure}

We follow the same procedure as in the case of TIM (given in \ref{appA}) to show that the optimal value of $\tau_k$ is the same for all system sizes of LTIM as well. For a two spin system, the probability of the system to be in the ground state after adiabatic evolution at {\bf A} is given by $\cos(B_z \tau_k)^4 \cos(2J \tau_k)^2 + \sin(2 J \tau_k)^2 \sin(B_z \tau_k)^4$ and that for $L=4$ is given by
$( \cos(B_z \tau_k)^4 (\cos(J \tau_k)^4 + \sin(J \tau_k)^4) - 2 \sin( J \tau_k)^2 \cos(J \tau_k)^2  \sin(B_z \tau_k)^4 )^2$. The maxima for these functions when $B_z$ and $J$ is set to unity, occur at $\tau_k^{opt} = n \pi$ with $n = 0, 1, 2 ..$ which gives the $\tau_k^{opt}$ values. This is verified in figure (\ref{fig_ltim_ee}) where we plot $\mathcal{E}_{A^{\prime}}$ and $\mathcal{E}_{A}$ as a function of $\tau_k$ for different system sizes. It is observed that even in the diabatic limit ($\tau_2 =0.1$), the minimum $\mathcal{E}_{A}$ occurs at the same $\tau_k$ for both $L=2$ and $L=10$, implying that $\tau_k^{opt}$ indeed is system size independent.

\section*{References}
%\bibliography{ref}

\begin{thebibliography}{10}

\bibitem{Mukherjee_2021}
Victor Mukherjee and Uma Divakaran.
\newblock Many-body quantum thermal machines.
\newblock {\em Journal of Physics: Condensed Matter}, 33(45):454001, aug 2021.

\bibitem{bhattacharjee21quantum}
Sourav Bhattacharjee and Amit Dutta.
\newblock Quantum thermal machines and batteries.
\newblock {\em The European Physical Journal B}, 94(12):239, Dec 2021.

\bibitem{myers22quantum}
Nathan~M. Myers, Obinna Abah, and Sebastian Deffner.
\newblock Quantum thermodynamic devices: From theoretical proposals to
  experimental reality.
\newblock {\em AVS Quantum Science}, 4(2):027101, 2022.

\bibitem{e23030353}
Jin-Fu Chen, Ying Li, and Hui Dong.
\newblock Simulating finite-time isothermal processes with superconducting
  quantum circuits.
\newblock {\em Entropy}, 23(3), 2021.

\bibitem{Abah_2016}
Obinna Abah and Eric Lutz.
\newblock Optimal performance of a quantum otto refrigerator.
\newblock {\em Europhysics Letters}, 113(6):60002, apr 2016.

\bibitem{Maslennikov17}
Gleb Maslennikov, Shiqian Ding, Roland Habl{\"u}tzel, Jaren Gan, Alexandre
  Roulet, Stefan Nimmrichter, Jibo Dai, Valerio Scarani, and Dzmitry
  Matsukevich.
\newblock Quantum absorption refrigerator with trapped ions.
\newblock {\em Nature Communications}, 10(1):202, 2019.

\bibitem{PhysRevA.97.022106}
Thao~P. Le, Jesper Levinsen, Kavan Modi, Meera~M. Parish, and Felix~A. Pollock.
\newblock Spin-chain model of a many-body quantum battery.
\newblock {\em Phys. Rev. A}, 97:022106, Feb 2018.

\bibitem{PhysRevE.106.024110}
Nikhil Gupt, Srijan Bhattacharyya, Bikash Das, Subhadeep Datta, Victor
  Mukherjee, and Arnab Ghosh.
\newblock Floquet quantum thermal transistor.
\newblock {\em Phys. Rev. E}, 106:024110, Aug 2022.

\bibitem{gemmer2009quantum}
Jochen Gemmer, Mathias Michel, and G{\"u}nter Mahler.
\newblock {\em Quantum thermodynamics: Emergence of thermodynamic behavior
  within composite quantum systems}, volume 784.
\newblock Springer, 2009.

\bibitem{PhysRevE.76.031105}
H.~T. Quan, Yu-xi Liu, C.~P. Sun, and Franco Nori.
\newblock Quantum thermodynamic cycles and quantum heat engines.
\newblock {\em Phys. Rev. E}, 76:031105, Sep 2007.

\bibitem{rossnagel16a}
Johannes Ro{\ss}nagel, Samuel~T. Dawkins, Karl~N. Tolazzi, Obinna Abah, Eric
  Lutz, Ferdinand Schmidt-Kaler, and Kilian Singer.
\newblock A single-atom heat engine.
\newblock {\em Science}, 352(6283):325--329, 2016.

\bibitem{PhysRevLett.123.080602}
D.~von Lindenfels, O.~Gr\"ab, C.~T. Schmiegelow, V.~Kaushal, J.~Schulz, Mark~T.
  Mitchison, John Goold, F.~Schmidt-Kaler, and U.~G. Poschinger.
\newblock Spin heat engine coupled to a harmonic-oscillator flywheel.
\newblock {\em Phys. Rev. Lett.}, 123:080602, Aug 2019.

\bibitem{schreiber15observation}
Michael Schreiber, Sean~S. Hodgman, Pranjal Bordia, Henrik~P. L{\"u}schen,
  Mark~H. Fischer, Ronen Vosk, Ehud Altman, Ulrich Schneider, and Immanuel
  Bloch.
\newblock Observation of many-body localization of interacting fermions in a
  quasirandom optical lattice.
\newblock {\em Science}, 349(6250):842--845, 2015.

\bibitem{Peterson18}
John P.~S. Peterson, Tiago~B. Batalh\~ao, Marcela Herrera, Alexandre~M. Souza,
  Roberto~S. Sarthour, Ivan~S. Oliveira, and Roberto~M. Serra.
\newblock Experimental characterization of a spin quantum heat engine.
\newblock {\em Phys. Rev. Lett.}, 123:240601, Dec 2019.

\bibitem{klatzow19experimental}
James Klatzow, Jonas~N. Becker, Patrick~M. Ledingham, Christian Weinzetl,
  Krzysztof~T. Kaczmarek, Dylan~J. Saunders, Joshua Nunn, Ian~A. Walmsley, Raam
  Uzdin, and Eilon Poem.
\newblock Experimental demonstration of quantum effects in the operation of
  microscopic heat engines.
\newblock {\em Phys. Rev. Lett.}, 122:110601, Mar 2019.

\bibitem{PhysRevE.61.4774}
Tova Feldmann and Ronnie Kosloff.
\newblock Performance of discrete heat engines and heat pumps in finite time.
\newblock {\em Phys. Rev. E}, 61:4774--4790, May 2000.

\bibitem{PhysRevE.87.012140}
D.~Gelbwaser-Klimovsky, R.~Alicki, and G.~Kurizki.
\newblock Minimal universal quantum heat machine.
\newblock {\em Phys. Rev. E}, 87:012140, Jan 2013.

\bibitem{PhysRevA.99.062103}
Patrice~A. Camati, Jonas F.~G. Santos, and Roberto~M. Serra.
\newblock Coherence effects in the performance of the quantum otto heat engine.
\newblock {\em Phys. Rev. A}, 99:062103, Jun 2019.

\bibitem{PhysRevE.83.031135}
George Thomas and Ramandeep~S. Johal.
\newblock Coupled quantum otto cycle.
\newblock {\em Phys. Rev. E}, 83:031135, Mar 2011.

\bibitem{Campisi_2015}
Michele Campisi, Jukka Pekola, and Rosario Fazio.
\newblock Nonequilibrium fluctuations in quantum heat engines: theory, example,
  and possible solid state experiments.
\newblock {\em New Journal of Physics}, 17(3):035012, mar 2015.

\bibitem{Cakmak_2023}
Sel{\c{c}}uk {\c{C}}akmak and H~R~Rastegar Sedehi.
\newblock Construction of a quantum stirling engine cycle tuned by
  dynamic-angle spinning.
\newblock {\em Physica Scripta}, 98(10):105921, sep 2023.

\bibitem{Cakmak2021}
Sel{\c{c}}uk {\c{C}}akmak and Ferdi Altintas.
\newblock Different constructions and optimization of the irreversible quantum
  carnot cycle.
\newblock {\em The European Physical Journal Plus}, 136(4):13, april 2021.

\bibitem{e19040136}
Ronnie Kosloff and Yair Rezek.
\newblock The quantum harmonic otto cycle.
\newblock {\em Entropy}, 19(4), 2017.

\bibitem{Rezek_2006}
Yair Rezek and Ronnie Kosloff.
\newblock Irreversible performance of a quantum harmonic heat engine.
\newblock {\em New Journal of Physics}, 8(5):83, may 2006.

\bibitem{Niedenzu_2018}
Wolfgang Niedenzu and Gershon Kurizki.
\newblock Cooperative many-body enhancement of quantum thermal machine power.
\newblock {\em New Journal of Physics}, 20(11):113038, nov 2018.

\bibitem{PhysRevLett.124.210603}
Gentaro Watanabe, B.~Prasanna Venkatesh, Peter Talkner, Myung-Joong Hwang, and
  Adolfo del Campo.
\newblock Quantum statistical enhancement of the collective performance of
  multiple bosonic engines.
\newblock {\em Phys. Rev. Lett.}, 124:210603, May 2020.

\bibitem{Jaramillo16}
J~Jaramillo, M~Beau, and A~del Campo.
\newblock Quantum supremacy of many-particle thermal machines.
\newblock {\em New J. Phys.}, 18(7):075019, 2016.

\bibitem{PhysRevLett.131.210401}
Alberto Rolandi, Paolo Abiuso, and Mart\'{\i} Perarnau-Llobet.
\newblock Collective advantages in finite-time thermodynamics.
\newblock {\em Phys. Rev. Lett.}, 131:210401, Nov 2023.

\bibitem{Hardal15}
Ali {\"U}.~C. Hardal and {\"O}zg{\"u}r~E. M{\"u}stecaplıo{\u g}lu.
\newblock Superradiant quantum heat engine.
\newblock {\em Scientific Reports}, 5:12953, 2015.

\bibitem{PhysRevB.99.024203}
Nicole Yunger~Halpern, Christopher~David White, Sarang Gopalakrishnan, and Gil
  Refael.
\newblock Quantum engine based on many-body localization.
\newblock {\em Phys. Rev. B}, 99:024203, Jan 2019.

\bibitem{campisi2016power}
M.~Campisi and R.~Fazio.
\newblock The power of a critical heat engine.
\newblock {\em Nat. Commun.}, 7:11895, 2016.

\bibitem{Fogarty_2021}
Thomás Fogarty and Thomas Busch.
\newblock A many-body heat engine at criticality.
\newblock {\em Quantum Science and Technology}, 6(1):015003, nov 2020.

\bibitem{10.1088/1367-2630/ac963b}
Giulia Piccitto, Michele Campisi, and Davide Rossini.
\newblock The ising critical quantum otto engine.
\newblock {\em New Journal of Physics}, 2022.

\bibitem{PhysRevE.96.022143}
Yu-Han Ma, Shan-He Su, and Chang-Pu Sun.
\newblock Quantum thermodynamic cycle with quantum phase transition.
\newblock {\em Phys. Rev. E}, 96:022143, Aug 2017.

\bibitem{PhysRevResearch.2.043247}
B.~S Revathy, Victor Mukherjee, Uma Divakaran, and Adolfo del Campo.
\newblock Universal finite-time thermodynamics of many-body quantum machines
  from kibble-zurek scaling.
\newblock {\em Phys. Rev. Research}, 2:043247, Nov 2020.

\bibitem{PhysRevLett.109.115703}
Adolfo del Campo, Marek~M. Rams, and Wojciech~H. Zurek.
\newblock Assisted finite-rate adiabatic passage across a quantum critical
  point: Exact solution for the quantum ising model.
\newblock {\em Phys. Rev. Lett.}, 109:115703, Sep 2012.

\bibitem{kolodrubetz17geometry}
Michael Kolodrubetz, Dries Sels, Pankaj Mehta, and Anatoli Polkovnikov.
\newblock Geometry and non-adiabatic response in quantum and classical systems.
\newblock {\em Physics Reports}, 697:1--87, 2017.

\bibitem{RevModPhys.91.045001}
D.~Gu\'ery-Odelin, A.~Ruschhaupt, A.~Kiely, E.~Torrontegui,
  S.~Mart\'{\i}nez-Garaot, and J.~G. Muga.
\newblock Shortcuts to adiabaticity: Concepts, methods, and applications.
\newblock {\em Rev. Mod. Phys.}, 91:045001, Oct 2019.

\bibitem{delcampo14}
A.~del Campo, J.~Goold, and M.~Paternostro.
\newblock More bang for your buck: Super-adiabatic quantum engines.
\newblock {\em Sci. Rep.}, 4:6208, 2014.

\bibitem{Deng13}
Jiawen Deng, Qing-hai Wang, Zhihao Liu, Peter H\"anggi, and Jiangbin Gong.
\newblock Boosting work characteristics and overall heat-engine performance via
  shortcuts to adiabaticity: Quantum and classical systems.
\newblock {\em Phys. Rev. E}, 88:062122, Dec 2013.

\bibitem{Beau16}
Mathieu Beau, Juan Jaramillo, and Adolfo del Campo.
\newblock Scaling-up quantum heat engines efficiently via shortcuts to
  adiabaticity.
\newblock {\em Entropy}, 18:168, 2016.

\bibitem{sels17minimizing}
Dries Sels and Anatoli Polkovnikov.
\newblock Minimizing irreversible losses in quantum systems by local
  counterdiabatic driving.
\newblock {\em Proceedings of the National Academy of Sciences},
  114(20):E3909--E3916, 2017.

\bibitem{PhysRevE.98.032121}
Obinna Abah and Eric Lutz.
\newblock Performance of shortcut-to-adiabaticity quantum engines.
\newblock {\em Phys. Rev. E}, 98:032121, Sep 2018.

\bibitem{delCampo2018}
Adolfo del Campo, Aur{\'e}lia Chenu, Shujin Deng, and Haibin Wu.
\newblock {\em Friction-Free Quantum Machines}, pages 127--148.
\newblock Springer International Publishing, Cham, 2018.

\bibitem{PhysRevResearch.2.023145}
Andreas Hartmann, Victor Mukherjee, Wolfgang Niedenzu, and Wolfgang Lechner.
\newblock Many-body quantum heat engines with shortcuts to adiabaticity.
\newblock {\em Phys. Rev. Research}, 2:023145, May 2020.

\bibitem{Diao_2018}
Pengpeng Diao, Shujin Deng, Fang Li, Shi Yu, Aurélia Chenu, Adolfo del Campo,
  and Haibin Wu.
\newblock Shortcuts to adiabaticity in fermi gases.
\newblock {\em New Journal of Physics}, 20(10):105004, oct 2018.

\bibitem{doi:10.1126/sciadv.aar5909}
Shujin Deng, Aurélia Chenu, Pengpeng Diao, Fang Li, Shi Yu, Ivan Coulamy,
  Adolfo del Campo, and Haibin Wu.
\newblock Superadiabatic quantum friction suppression in finite-time
  thermodynamics.
\newblock {\em Science Advances}, 4(4):eaar5909, 2018.

\bibitem{chen2019interaction}
Yang-Yang Chen, Gentaro Watanabe, Yi-Cong Yu, Xi-Wen Guan, and Adolfo del
  Campo.
\newblock An interaction-driven many-particle quantum heat engine and its
  universal behavior.
\newblock {\em npj Quantum Information}, 5(1):88, 2019.

\bibitem{PhysRevResearch.5.013088}
Mohamed Boubakour, Thom\'as Fogarty, and Thomas Busch.
\newblock Interaction-enhanced quantum heat engine.
\newblock {\em Phys. Rev. Res.}, 5:013088, Feb 2023.

\bibitem{revathybatheng}
B.~S Revathy, Victor Mukherjee, and Uma Divakaran.
\newblock Bath engineering enhanced quantum critical engines.
\newblock {\em Entropy}, 24(10):1458, 2022.

\bibitem{PhysRevA.82.063422}
Dionisis Stefanatos, Justin Ruths, and Jr-Shin Li.
\newblock Frictionless atom cooling in harmonic traps: A time-optimal approach.
\newblock {\em Phys. Rev. A}, 82:063422, Dec 2010.

\bibitem{doi:10.1021/acs.jpca.5b06090}
Yang-Yang Cui, Xi~Chen, and J.~G. Muga.
\newblock Transient particle energies in shortcuts to adiabatic expansions of
  harmonic traps.
\newblock {\em The Journal of Physical Chemistry A}, 120(19):2962--2969, 2016.

\bibitem{lieb61two}
Elliott Lieb, Theodore Schultz, and Daniel Mattis.
\newblock Two soluble models of an antiferromagnetic chain.
\newblock {\em Annals of Physics}, 16(3):407 -- 466, 1961.

\bibitem{pfeuty70the}
Pierre Pfeuty.
\newblock The one-dimensional ising model with a transverse field.
\newblock {\em Annals of Physics}, 57(1):79 -- 90, 1970.

\bibitem{bunder99effect}
J.~E. Bunder and Ross~H. McKenzie.
\newblock Effect of disorder on quantum phase transitions in anisotropic xy
  spin chains in a transverse field.
\newblock {\em Phys. Rev. B}, 60:344--358, Jul 1999.

\bibitem{sachdev_2011}
Subir Sachdev.
\newblock {\em Quantum Phase Transitions}.
\newblock Cambridge University Press, 2 edition, 2011.

\bibitem{dutta15quantum}
A.~Dutta, G.~Aeppli, B.~K. Chakrabarti, U.~Divakaran, T.~F. Rosenbaum, and
  D.~Sen.
\newblock {\em Quantum phase transitions in transverse field spin models: from
  statistical physics to quantum information}.
\newblock Cambridge University Press, Cambridge, 2015.

\bibitem{RevModPhys.83.863}
Anatoli Polkovnikov, Krishnendu Sengupta, Alessandro Silva, and Mukund
  Vengalattore.
\newblock Colloquium: Nonequilibrium dynamics of closed interacting quantum
  systems.
\newblock {\em Rev. Mod. Phys.}, 83:863--883, Aug 2011.

\bibitem{PhysRevB.92.104306}
Shraddha Sharma, Sei Suzuki, and Amit Dutta.
\newblock Quenches and dynamical phase transitions in a nonintegrable quantum
  ising model.
\newblock {\em Phys. Rev. B}, 92:104306, Sep 2015.

\bibitem{PhysRevE.99.012122}
O.~F. de~Alcantara Bonfim, B.~Boechat, and J.~Florencio.
\newblock Ground-state properties of the one-dimensional transverse ising model
  in a longitudinal magnetic field.
\newblock {\em Phys. Rev. E}, 99:012122, Jan 2019.

\end{thebibliography}
%\bibliographystyle{unsrt}

\end{document}